\documentclass[aps,prl,twocolumn,superscriptaddress,draft,showpacs,intlimits,amsmath,amssymb,floats]{revtex4}
\usepackage{bm}
\usepackage[final]{graphicx}
\usepackage{epsfig}
\usepackage{color}
\definecolor{DarkBlue}{rgb}{0.1,0.1,0.5}
\definecolor{Red}{rgb}{0.9,0.0,0.1}
\definecolor{Green}{rgb}{0.0,0.99,0.0}

\begin{document}
\title{Unraveling  the Nature of Charge Excitations in La$_2$CuO$_4$ with Momentum-Resolved Cu $K$-edge Resonant Inelastic X-ray Scattering}
\date{\today}
\author{C.-C. Chen}
\affiliation{Stanford Institute for Materials and Energy Science, SLAC National Accelerator Laboratory, Menlo Park, California 94025, USA}
\affiliation{Department of Physics, Stanford University, Stanford, California 94305, USA}
\author{B. Moritz}
\affiliation{Stanford Institute for Materials and Energy Science, SLAC National Accelerator Laboratory, Menlo Park, California 94025, USA}
\affiliation{Department of Physics and Astrophysics, University of North Dakota, Grand Forks, North Dakota 58202, USA}
\author{F. Vernay}
\affiliation{LAMPS, Universit\'{e} de Perpignan Via Domitia, F-66860 Perpignan, France}
\author{J. N. Hancock}
\affiliation{Stanford Synchrotron Radiation Laboratory, Stanford, California 94309, USA}
\affiliation{D\'{e}partment de Physique de la Mati\`{e}re Condens\'{e}e, 24 quai Ernest-Ansermet, Universit\'{e} de Gen\`{e}ve, Gen\`{e}ve, CH-1211, Switzerland}
\author{S. Johnston}
\affiliation{Stanford Institute for Materials and Energy Science, SLAC National Accelerator Laboratory, Menlo Park, California 94025, USA}
\affiliation{Department of Physics and Astronomy, University of Waterloo, Waterloo, Ontario N2L 3G1, Canada}
\author{C. J. Jia}
\affiliation{Stanford Institute for Materials and Energy Science, SLAC National Accelerator Laboratory, Menlo Park, California 94025, USA}
\affiliation{Department of Applied Physics, Stanford University, Stanford, California 94305, USA}
\author{G. Chabot-Couture}
\affiliation{Department of Applied Physics, Stanford University, Stanford, California 94305, USA}
\author{M. Greven}
\affiliation{Department of Applied Physics, Stanford University, Stanford, California 94305, USA} 
\affiliation{School of Physics and Astronomy, University of Minnesota, Minneapolis, Minnesota 55455, USA}
\author{I. Elfimov}
\affiliation{Department of Physics and Astronomy, University of British Columbia, Vancouver, British Columbia V6T 1Z1, Canada}  
\author{G. A. Sawatzky}
\affiliation{Department of Physics and Astronomy, University of British Columbia, Vancouver, British Columbia V6T 1Z1, Canada}  
\author{T. P. Devereaux}
\affiliation{Stanford Institute for Materials and Energy Science, SLAC National Accelerator Laboratory, Menlo Park, California 94025, USA}

\begin{abstract}
Results of model calculations using exact diagonalization reveal the orbital character of states associated with different Raman loss peaks in Cu $K$-edge resonant inelastic X-ray scattering (RIXS) from La$_{2}$CuO$_{4}$. The model includes electronic orbitals necessary to highlight non-local Zhang-Rice singlet, charge transfer and $d$-$d$ excitations, as well as states with apical oxygen 2$p_z$ character. The dispersion of these excitations is discussed with prospects for resonant final state wave-function mapping. A good agreement with experiments emphasizes the substantial multi-orbital character of RIXS profiles in the energy transfer range 1-6 eV. 
\end{abstract}
\maketitle

Understanding the nature of charge excitations in correlated systems remains a challenge even after more than two decades of intense study. Recent experiments in the cuprate high-$T_c$ superconductors highlight the potential importance of additional orbital degrees of freedom involving the apex oxygens.~\cite{Greven_tsb} Moreover, a correlation between the apex oxygen site and $T_c$ has been made in different contexts.~\cite{Pavarini} However, to date no experiment has been able to directly test the apical character of the low energy states in the cuprates. Nonetheless, these results suggest a re-examination of the canonical view that the physics in cuprates is solely governed by electrons confined to the in-plane copper-oxygen orbitals.

One paradigmatic material is La$_2$CuO$_4$, on which many spectroscopic techniques have been used to elucidate various aspects of the underlying physics.~\cite{RMP_techniques} Among these techniques, resonant inelastic x-ray scattering (RIXS) \cite{RIXS-RMP} probes charge excitations in a momentum-resolved way. A net momentum $\Delta\mathbf{Q}$ is transferred during the RIXS process, and the associated photon wavelength allows access to the entire Brillouin zone (BZ). By tuning the x-ray energy to proper absorption edges, RIXS provides material-specific information of various many-body excitations, \textit{e.g.} magnon, $d$-$d$, and charge transfer (CT) excitations, in strongly correlated materials.~\cite{RIXS-RMP, YJK_MO, YJKim2008, Hill, Greven, Jason2008}

Cu $K$-edge ($1s$-$4p$ core-level excitation) RIXS measurements in the cuprates reveal a number of spectral peaks in the energy-loss range 1-6 eV,~\cite{YJK_MO, YJKim2008,Hill, Greven, Jason2008} with a most prominent peak $\sim$ 4-6 eV and another less prominent peak $\sim$2 eV at $\Delta\mathbf{Q}=(0,0)$ (the $\Gamma$ point or zone center). The former has been attributed to a local CT of a hole from Cu to its neighboring oxygen ligands, and the latter has been attributed to excitation from the lower Hubbard band across the band gap, leading to a CT excitation according to the Zaanen-Sawatzky-Allen scheme.~\cite{ZAS_classification} These and other excitations have been studied carefully in effective low energy Hamiltonians.~\cite{Maekawa} However, these down-folded approaches fail to fully account for the impact of many intertwined orbitals on RIXS profiles.

In this letter, a ``two-step" exact diagonalization (ED) algorithm facilitates the characterization of the Cu $K$-edge RIXS spectra in La$_{2}$CuO$_{4}$ in the energy range 1-6 eV with various $\Delta\mathbf{Q}$. These excitations is shown to be rich in orbital character and not described by the dispersion of a single Mott gap excitation. The $d$-$d$, apical oxygen, and CT excitations are all found in the studied energy loss range, having intensities that vary with $\Delta\mathbf{Q}$. The method consists of (1) a matrix diagonalization of the model Hamiltonian to obtain the ground state and final excited states, and (2) a series of diagonalizations to obtain intermediate states with the core-hole interaction. We use the parallel version of the ARPACK libraries, based on iterative Arnoldi methods,~\cite{ARPACK} to calculate the energy-eigenvalues and orthonormalized eigenfunctions.

We focused on quasi-2D copper-oxygen clusters with periodic boundary conditions. The Cu$_4$O$_x$ cluster allows investigations of three distinct $\Delta\mathbf{Q}$-points: $(0,0)$, $(\pi,0)$ and $(\pi,\pi)$, which is enough to demonstrate the qualitative differences in the momentum-dependent RIXS spectra. We use a multi-orbital Hubbard Hamiltonian including Cu $3d$ $e_g$ and bonding O $2p$ orbitals, with the Hund's coupling and direct exchange interaction to account for multiplet splitting.~\cite{DagottoCMR, H_parameters} Explicit photon-polarization dependence is introduced by convolving the RIXS spectrum with a Cu $4p$ DOS obtained from density functional theory (DFT) calculation.~\cite{Vernay2008}

\begin{figure}[t]
\includegraphics[width=\columnwidth]{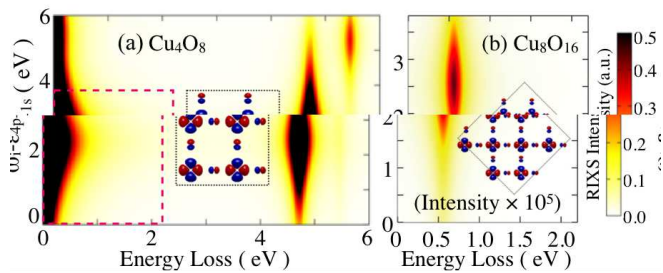}
\caption{(Color online) Theoretical RIXS spectra at the $\Gamma$ point for (a) planar Cu$_4$O$_8$, and (b) planar Cu$_8$O$_{16}$ cluster (with the elastic resonance removed). The red dashed rectangle in (a) indicates the energy range of the RIXS plot in (b) (intensity $\times 10^5$). The spectra are broadened with a 0.1 eV Lorentzian in energy loss and a core hole lifetime of 1.0 eV.
}\label{fig1}
\end{figure}

Results at the $\Gamma$ point are displayed in Figs. 1 and 2, where we show progressively more spectral complexity by including more orbitals. Retaining only Cu 3$d_{x^2-y^2}$ and its hybridized planar O 2$p_{x,y}$ orbitals [Cu$_4$O$_8$] yields Fig. 1(a). Besides the elastic feature, there is only a single prominent peak at energy transfer $\sim$ 4.8 eV. This peak stems from a local CT excitation of a Cu $d^9$ hole onto its surrounding ligand oxygens with $d_{x^2-y^2}$ symmetry, creating a local molecular orbital $d^{10}\underline{L}$. This energy loss is directly controlled by planar charge transfer energy $\Delta=\epsilon_p-\epsilon_d$, and is further increased by a level repulsion with the ground state via $p$-$d$ hybridization. The inclusion of an excitonic interaction between the Cu 3$d$ electron and O 2$p$ hole would tend to decrease the excitation energy. However, for the ligand state of $d_{x^2-y^2}$ symmetry this effective electron-hole interaction is small.~\cite{Varma, Higashiya}

We note that while the energy gap revealed by the calculated hole addition/removal spectra is $\sim$1.7 eV,~\cite{GAP} there is no RIXS resonance at this energy at the $\Gamma$ point associated with a Zhang-Rice singlet (ZRS) CT excitation. Similar calculations have been performed for a planar Cu$_8$O$_{16}$ cluster up to energy loss 2.2 eV. As shown in Fig. 1(b), a weak 0.56 eV bi-magnon peak is found, but a final state resonance $\sim$2 eV remains absent.

\begin{figure}[t!]
\includegraphics[width=\columnwidth]{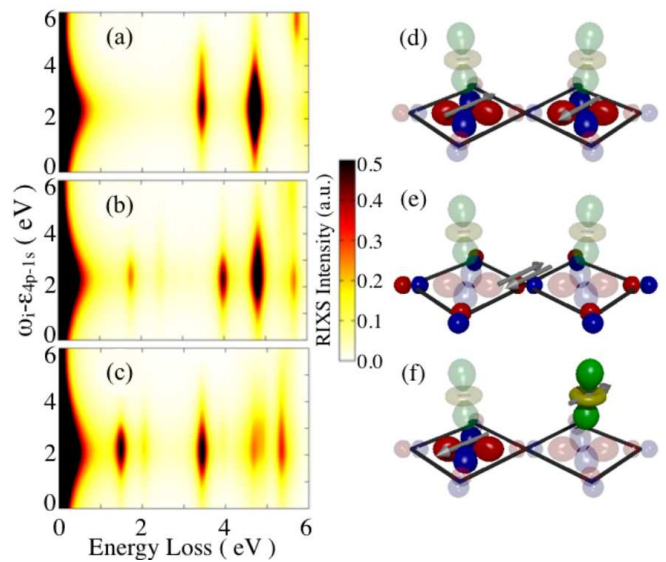}
\caption{
(Color online) Theoretical RIXS spectra at the $\Gamma$ point for (a) octahedral Cu$_4$O$_{16}$ with apical O, (b) planar Cu$^*_4$O$_8$ (with Cu 3$d_{3z^2-r^2})$, and (c) octahedral Cu$^*_4$O$_{16}$ (with both Cu 3$d_{3z^2-r^2}$ and apical O). (d)-(f) Schematic of the non-local inter-site $d$-$d$ excitation pathway for the results in (b) and (c): (d) $d^9$ ground state configuration, (e) intermediate ligand occupation on both plaquettes, and (f) final state configuration highlighting the inter-site $d$-$d$ excitation.}\label{non_local_dd}
\end{figure}

As shown previously,~\cite{ZR} a non-local ZRS CT excitation is well-defined away from the $\Gamma$ point (especially along the magnetic BZ boundary). Although the 2 eV ZRS feature is absent in the hole addition spectra at the $\Gamma$ point measured at half-filling by angle-resolved photoemission due to complete cancellation between $2p_{x,y}$-$3d_{x^2-y^2}$ hybridizations,~\cite{Pothuizen, QY2008} its absence in RIXS can be more subtle. Unlike ARPES, in $K$-edge RIXS the hole number is conserved, and a $d^{9}\underline{L}$ is created together with a $d^{10}$ via charge transfer. Certain configurations, where the momentum of a ZRS and a $d^{10}$ hole sum to zero, may admit a final state resonance associated with ZRS at the $\Gamma$ point. Therefore, the absence of a 2 eV feature may be an effect of confinement on the $d^{9}\underline{L}$ and $d^{10}$ and thus an artifact of the small cluster size. To provide a definitive answer regarding the existence of a 2eV feature in RIXS, it is essential to go beyond the Cu$_8$O$_{16}$ calculation.

We show results at the $\Gamma$ point that include apical oxygen $2p_z$, copper 3$d_{3z^2-r^2}$, and both orbitals in Fig. 2 (a)-(c), respectively. Apical oxygens [Cu$_4$O$_{16}$ in Fig. 2(a)] yield an additional peak compared to Fig. 1 at an energy loss $\sim$ 3.5 eV controlled by the apical CT energy, attributed to a nonlocal symmetry allowed CT excitation of the Cu $d^9$ hole onto inter-plaquette apex oxygens. Although the CT energies $\Delta$ are similar for planar and apical oxygen orbitals, this apical excitation lies below the $d^{10}\underline L$ due to a small level repulsion with the ground state, stemming from a lack of $2p_z$-$3d_{x^2-y^2}$ hybridization. 

Inclusion of the Cu 3$d_{3z^2-r^2}$ orbital (Figs. 2(b) [Cu$_4^*$O$_8$] and 2(c) [Cu$_4^*$O$_{16}$]) produces peaks at lower energies. These low energy peaks, present only with the $d_{3z^2-r^2}$ orbital, are of $d$-$d$ character. An intra-site $d$-$d$ excitation, where a Cu $3d_{3z^2-r^2}$ hole is left behind after $4p$-$1s$ recombination on the same Cu atom, is symmetry-forbidden.~\cite{Vernay2008,Jason2008} However, non-local inter-site $d$-$d$ excitation, where a Cu $3d_{3z^2-r^2}$ hole is excited from its neighboring Cu $3d_{x^2-y^2}$ orbitals, is symmetry-allowed [see Fig. \ref{non_local_dd}(d)-(f)]. The resonant features at loss energies $\sim$ 1.5 and 2.0 eV are characterized by single and double nonlocal $d$-$d$ excitations. While the bare crystal-field-splitting (CFS) between the two $e_g$ orbitals in the calculation is only 0.2 eV, the de-localization between the Cu 3$d_{x^2-y^2}$ hole and the ligand ($\sim 1.39$ eV)~\cite{MAV1993} causes the peak to appear at higher energies, essentially the same effect as the covalent contribution to the ligand field splitting.~\cite{Fujii} More peaks emerge around the $d^{10}\underline L$ molecular orbital excitation which contain different mixings of Cu $3d_{x^2-y^2}, 3d_{3z^2-r^2}$ and O $2p_{x,y}, 2p_z$ orbitals. Moreover, Figs. 2(b) [Cu$^*_4$O$_8$, relevant for Sr$_2$CuO$_2$Cl$_2$ and Nd$_2$CuO$_4$] and 2(c) indicate that the presence of apex oxygens shifts the RIXS peaks to lower energy, due to kinetic energy stabilization of states with $3d_{3z^2-r^2}$ character. Taken as a whole, these results demonstrate that the orbital character of charge excitations accessible via RIXS becomes increasingly more complex and material dependent at higher energy transfer.

\begin{figure}[t]
\includegraphics[width=\columnwidth]{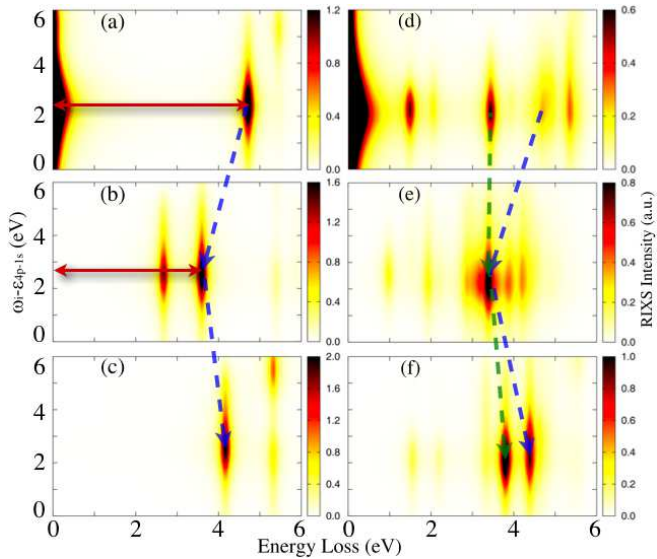}
\caption{
(Color online) Theoretical RIXS spectra for Cu$_4$O$_{8}$ at $\Delta \mathbf{Q} =$ (a) $(0, 0)$, (b) $(\pi, 0)$, and (c) $(\pi, \pi)$.  (d), (e), and (f) represent the same momentum points for Cu$_4^*$O$_{16}$. The dispersions of local molecular orbital and inter-plaquette apical CT excitations are indicated respectively by the blue and green dashed lines. 
}\label{momentum}
\end{figure}

In Fig. \ref{momentum} we describe the dispersion of various RIXS excitations. First, for Cu$_4$O$_8$ [Figs. \ref{momentum} (a)-(c)] the prominent molecular orbital excitation energy softens from $(0,0)$ to $(\pi,0)$, and increases again at $(\pi,\pi)$ [follow the dashed line in Fig. \ref{momentum}(a)], yet remains lower than its energy at the $\Gamma$ point, indicating an indirect energy gap. This behavior agrees with experimental observations.~\cite{YJK_MO} At finite momentum transfer $\Delta\mathbf{Q}=(\pi,0)$, $d^{10}\underline{L}$ excitation propagates more easily because of its associated ligand symmetry. As shown previously,~\cite{ZR} a non-local ZRS CT excitation is well-defined along the magnetic BZ boundary and does not suffer from the potential impact of the small cluster size. This is confirmed in Figs. \ref{momentum} (b) and (e), where a smaller non-local ZRS CT peak emerges at $\sim$ 2.8 eV at $\Delta \mathbf{Q}=(\pi,0)$, $\sim$800 meV higher in energy than that of the anticipated ZRS at the $\Gamma$ point.

The Cu$_4^*$O$_{16}$ dispersion profile [Fig. \ref{momentum}(d)-(f)] can be understood with help from the Cu$_4$O$_8$ analysis. The dominant peak at $\Delta \mathbf{Q} =(\pi,0)$ at $\sim$ 3.5 eV results from the local molecular orbital excitation. On the other hand, the $\sim$ 3.5 eV inter-plaquette apical CT excitation at the $\Gamma$ point has little mixing with the ground state, and therefore only weakly disperses. The RIXS data, focusing on the dispersion of the excited states and their characters, thereby provide a vehicle to probe the apical character of the many-body states. The cluster calculations give only a few percent apical weight in the ground state,~\cite{CTChen_XAS} implying that phenomena associated with apical ground state character could be difficult to detect.

\begin{figure}[t!]
\includegraphics[width=\columnwidth]{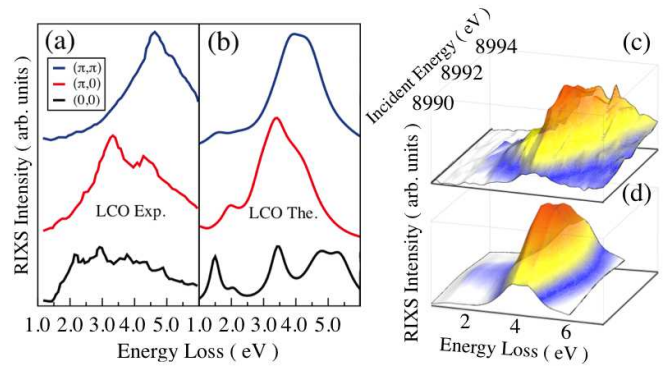}
\caption{
(Color online) (a) Experimental~\cite{YJKim2008} and (b) theoretical RIXS spectra for 
different $\Delta \mathbf{Q}$ at incident photon energy 8992.5 eV. The elastic part for $\Delta \mathbf{Q}=(0,0)$ is removed. The spectra in (b) are broadened with an energy-dependent Lorentzian increasing linearly with energy loss from 0.1 to 0.4 eV, 0.2 to 0.5 eV, and 0.3 to 0.6 eV, for $\Delta \mathbf{Q}=(0,0)$, $(\pi,0)$, and $(\pi,\pi)$, respectively. (c) Experimental and (d) theoretical RIXS spectra convolved with $4p_z$ DOS at $\Delta \mathbf{Q}=(\pi,\pi)$.
}\label{YJKim_expt}
\end{figure}

Figure \ref{YJKim_expt}(a) displays cuts of experimental RIXS spectra on La$_2$CuO$_4$ at three different $\Delta \mathbf{Q}$ points.~\cite{YJKim2008} The RIXS profile broadens into a single peak as $\Delta \mathbf{Q}$ increases. As higher energy states lie in a high density continua, we use energy-dependent broadening for different $\Delta \mathbf{Q}$. A generally good agreement with the experiments at finite $\Delta \mathbf{Q}$ results from this procedure, particularly for energies above 3 eV. However, the calculation does not obtain the broad intensity between energy loss $2\sim3$ eV at the $\Gamma$ point, and seems to overestimate the $d$-$d$ excitation intensity. Agreement may be improved at lower energies with larger cluster studies or smaller Hilbert space calculations with downfolded Hamiltonians, which would emphasize the ZRS CT with a likely suppression of $d$-$d$ spectral weight. We also remark that a nonlocal ZRS CT excitation with momentum $\Delta \mathbf{Q}$ may appear at the $\Gamma$ point with another excitation, such as a phonon or a magnon, carrying momentum $-\Delta \mathbf{Q}$.

Finally we demonstrate that incorporating 4$p_z$ DOS obtained from DFT produces an adequate position for the $K$-edge resonance and the main energy-loss feature. RIXS data for La$_2$CuO$_4$ from Ref.~\onlinecite{Greven} are replotted in Fig. \ref{YJKim_expt}(c). The spectra were collected at room-temperature in vertical scattering geometry, with $\hat\epsilon$ $\|$ $\hat{c}$ ($\Delta\mathbf{Q}$=(2.5,0.5,0)) and overall energy resolution set to $ \sim$ 0.3 eV. Experimental RIXS spectra have broad features, with a dominant peak at incident photon energy $\sim$ 8992 eV and energy loss $\sim$ 4 eV, agreeing with the calculation [Fig. \ref{YJKim_expt}(d)].

In summary, this paper describes a nontrivial momentum dependence of the RIXS process and the resulting multi-orbital character of the RIXS profile for La$_2$CuO$_4$. We demonstrate how the multi-orbital character of excited states may be mapped throughout the BZ, and give a reason why a dispersion of certain peaks may be associated with a change of character. Moreover, we show that subdominant character of the ground state may be revealed as apparent dispersion of a local excitation.

We also demonstrate how the presence of the apical oxygens can affect the ground state and the excitation spectrum. From the spectral weight and dispersion, one can confirm whether there is substantial apical character of the ground state. As the apical oxygen energy is one of the few parameters that changes across the high $T_c$ compounds, $K$-edge RIXS in the energy range 1-6 eV may offer a diagnostic tool to examine material dependence, and provide hints to the material-specific $T_c$.~\cite{two-orbital}

For the undoped cluster, the apical oxygen character, including that associated with a possible time-reversal symmetry breaking (TRSB) phase involving circulating current loops through the apical oxygens, has very small weight in the ground state, with the majority appearing in higher energy excited states. This is consistent with variational Quantum Monte Carlo and cluster $t$-$J$ results which found evidence for such a phase only for larger doped clusters.~\cite{CC_Phase}. RIXS in principle can shed light onto a putative TRSB phase by performing circularly polarized dichroism experiments at the corresponding ordering wave-vector. An extension of the current calculations to address this interesting point for doped multiband clusters is an area of future study.

\textbf{Acknowledgment - } The authors thank J. van den Brink, Y.-J. Kim, W.-S. Lee, J.-H. Chu, K. Ko, and L.-Q. Lee for discussions. This work is supported by the U.S. DOE under contract number DE-AC02-76SF00515, and DE-FG02-08ER46540 (CMSN). This research used resources of the NERSC, supported by DOE under Contract No. DE-AC02-05CH11231. CCC acknowledges support from NSC, Taiwan, under contract No. NSC-095-SAF-I-564-013-TMS. SJ acknowledges support from NSERC and SHARCNET. TPD, BM, FV, and SJ are grateful for the hospitality of PITP.

\end{document}